\documentstyle[epsfig,11pt]{article}
\begin{document}
\begin{flushright}
\begin{minipage}{6cm}
Preprint of Letter to the Editor\\ 
To be published in J. Phys. G\\
Version, \today\\
\end{minipage}
\end{flushright}
\begin{center}
\begin{Large}
{\bf Further search for a neutral boson \\ with a mass
around 9 MeV/c$^{2}$\\}
\end{Large}
\renewcommand{\thefootnote}{\fnsymbol{footnote}}
\vspace{0.1in}
F.W.N. de Boer$^{2,}$\footnote{E-mail address: fokke@nikhef.nl}, 
K. Bethge$^{1}$, H. Bokemeyer$^{3}$,
R. van Dantzig$^{2}$,\\ J. van Klinken$^{4}$, V. Mironov$^{5}$,
K.A. M\"{u}ller$^{1}$ and  K.E. Stiebing$^{1}$\\
\end{center}

\protect\begin{minipage}[t]{5.00in}
\begin{center}
\small{$^{1)}$ Institut f\"{u}r Kernphysik der Johann Wolfgang 
Goethe-Universit\"{a}t,\\ D-60486 Frankfurt am Main, Germany\\} 
\small{$^{2)}$ Nationaal Instituut voor Kernfysica en Hoge-Energie Fysica,\\ 
NL-1098 SJ Amsterdam, The Netherlands\\} 
\small{$^{3)}$ Gesellschaft f\"{u}r Schwerionenforschung, D-64291 Darmstadt, 
Germany\\} 
\small{$^{4)}$ Kernfysisch Versneller Instituut, 9747 AA Groningen,
The Netherlands\\}
\small{$^{5)}$ Laboratory of Particle Physics, Joint Institute for Nuclear 
Research,\\ 141980, Dubna, Moscow Region, Russia\\} 
\end{center}
\protect\end{minipage}

\vspace{2mm}
\begin{abstract}
Two dedicated experiments on internal pair conversion (IPC) of isoscalar M1
transitions were carried out 
in order to test a 9 MeV/c$^{2}$ $X$-boson scenario.
In the $^{7}$Li($p,e^{+}e^{-})^{8}$Be reaction at 1.1 MeV proton energy to the
predominantly T=0 level at 18.15 MeV, a significant deviation from IPC  
was observed at large pair correlation angles. 
In the $^{11}$B($d,ne^{+}e^{-})^{12}$C
reaction at 1.6 MeV, leading to the 12.71 MeV $1^{+}$ level with
pure T=0 character, an anomaly was observed at 9 MeV/c$^{2}$. 
The compatibility of the results with the scenario is discussed.
\end{abstract}

\vspace{2mm}
{\bf 1. Introduction}
\vspace{2mm}

A neutral boson can in principle be emitted in a nuclear 
transition within
the constraints of spin-parity and energy-momentum conservation. 
In magnetic transitions
such a boson should have a pseudoscalar ($0^{-}$) or an axial vector 
$(1^{+})$ character,
and in electric transitions a scalar $(0^{+})$ or a vector $(1^{-})$
character. 
When its mass is above 1.022 MeV, the signature of $X$-boson emission could be 
the two-body decay into $e^+e^-$ pairs 
superposed on conventional 
internal pair conversion \mbox{(IPC) \cite{rose}.} 
In a survey \cite{fokke2,fokke3} of data for IPC of transitions 
in several nuclei [4-8] with energies above 10 MeV, 
an anomaly with respect to IPC appears at an invariant mass of approximately 9
MeV/c$^{2}$.
Whereas (partial) isoscalar M1 transitions exhibit an excess 
of $e^{+}e^{-}$ pairs at large opening angles 
no deviations occur in isovector E1 and M1 transitions \cite{fokke1}.
Interpreting this anomaly in the perspective of a 
short-lived \mbox{(${\tau} {\leq} 10^{-10}$ s)}
neutral \mbox{$X$-boson} of \mbox{9 MeV/c$^{2}$ 
 \cite{fokke2,fokke3},} 
the deviations in 
isoscalar M1 transitions would indicate an isoscalar character 
for such a boson with spin-parity 0$^{-}$ or 1$^{+}$.

The branching ratio \cite{fokke2,fokke3} for emission of an elusive $X$-boson
with respect to \mbox{$\gamma$-emission} in  nuclear dipole transitions can be 
written as
\begin{equation}
     B_{X} = \frac{\alpha_{X}}{2{\alpha}} {\beta}^{3},
\end{equation}
where  $\alpha_{X}$ is the
the isospin dependent effective boson-nucleon coupling strength,
relative to $\tilde{\alpha} = 1.7 {\cdot} 10^{-6}$ (the axion-nucleon
coupling strength),
$\alpha$ the fine structure constant.
Measured values for $B_X$, ${\Gamma}_X$, ${\alpha}_X$ and $m_{X}$ for the different 
M1 transitions are listed in Table 1. Results from the previous studies
have been included.  The average $m_{X}$ is found to be 
($9.1 {\pm} 0.4$) MeV/c$^{2}$.

The investigations were triggered, about a decade ago,  
by the observation of  $e^+e^-$ pairs in an 
emulsion study \cite{elnadibadawy,fokkeemulsion1} at short 
distance from the interaction vertex, using $^{4}$He, $^{12}$C and $^{22}$Ne
projectiles at 4.5 GeV/A and $^{32}$S at 200 GeV/A. 
In a plot \cite{fokke2} 
of the invariant mass $m_X$ against the lifetime ${\tau}$, 
derived from the observed parameters of the pairs, a distinct cluster of 10 events 
with average 
invariant mass of
8.8 MeV/c$^{2}$ and lifetime of 6${\cdot}10^{-16}$ s was observed,
with a background of external pair conversion (EPC) \cite{borsellino}
from ${\gamma}$-rays mostly from ${\pi}^{o}$ decay. 

In the field of very high energy cosmic-ray interactions, 
possibly related results came from the the JACEE collaboration 
\cite{jacee1,jacee2}.
In emulsion chambers 
multiple electron pairs were observed in heavy particle decays, which 
could not be explained in terms
of known decay modes. 
However, the pairs were found to be consistent with the hypothesis
that some as yet unidentified, light neutral particles 
with masses less than 100 MeV/c$^{2}$ 
were emitted in bottom decay, which subsequently decayed into $e^+e^-$ pairs. 
Quite recently, revisiting the data, the observations
were reported \cite{jacee3,jacee4} 
to be not inconsistent with a 
light neutral boson with a mass of 9 MeV/c$^{2}$  
and lifetime of $1.4{\cdot}10^{-15}$ s. 

\begin{table}
\scalebox{0.85}{
\begin{tabular}{|cccccccccc|}
\hline
$^{A}$Z & E$_R$ & ${\Gamma}_R$&
 I$^{\pi}$,T& E$_{\gamma}$&  $B_{X}$ & ${\Gamma}_X$ & 
$\alpha_X$& $m_X$ & Refs.\\
 &MeV&eV& &MeV & & meV &$1.7{\cdot}10^{-6}$& MeV/c$^{2}$ & \\
\hline
$^{12}$C & 12.71 & 18.1 &$1^{+},0$ 
& 12.71   &   
$(7 \pm 3){\cdot}10^{-4}$  & $0.24 \pm 0.11$ & $18 {\pm} 7$ 
& $9.0 {\pm} 1.0$ &present\\
         &              &     &    
& 12.71 &   
$(1.6 \pm 0.7){\cdot}10^{-3}$  
& $0.56 {\pm} 0.25$ &  $38 {\pm} 17$ & $9.2 {\pm} 1.0$  & [5-7]\\ 
 & 15.11 & 43.6 &$1^{+},1$ &
 15.11& $\leq 4.6 {\cdot}10^{-5}$& $\leq 1.8$
&${\leq} 0.9$& -   
& [5-7]\\
$^{8}$Be & 17.64  & $10.7{\cdot}10^3$ & 1$^{+}$,1&  
 17.64  & 
$(1.1 \pm 0.3){\cdot}10^{-4}$ & $1.9 \pm 0.4$ & $1.5 \pm 0.4$ & 
$ 9 {\pm} 1 $ &
 [2-4]\\
 &  && 
   & 14.64 & $(8.5 \pm 2.6){\cdot}10^{-5}$ & $0.7 {\pm} 0.2$ & 
$1.5 \pm 0.4$ & $9 {\pm} 1$ & 
 [2-4]\\
&  18.15 &$138{\cdot}10^3$ & 1$^{+}$,0 
& 18.15 &  $\leq 4.1{\cdot}10^{-4}$  &  
${\leq} 1.2$ & ${\leq} 5.7$& - & present\\
 &   && & 
  15.15 &  $(5.8 {\pm} 2.2) {\cdot}10^{-4}$& 
$2.2 \pm 0.8$ &$10.5 \pm 4.5$ & $9.5 {\pm} 1.2$ & present\\
$^{4}$He & 21.0 & $850{\cdot}10^3$ & 0$^{-}$,0 & 
 M0   & $0^{-}{\rightarrow}0^{+}$,  $e^+e^-$    &
$ 74 {\pm} 30$ & $32 \pm 12$ & $8 \pm 2$ 
& [5-7]\\
\hline
\end{tabular}
}

\caption[Experimental]{\small {\em Experimental results 
for anomalous $e^+e^-$-emission interpreted in the light
of a short-lived
9 MeV/c$^{2}$ $X$-boson 
in six M1 transitions and an M0 transition. 
Listed are the nucleus, the energy and the width of the resonance E$_R$
and ${\Gamma}_R$, 
the (iso)spin-parity quantum numbers,
the transition energy $E_{\gamma}$, 
the $X$-branching ratio $B_X$ with respect to ${\gamma}$-emission,
the $X$-decay width ${\Gamma}_X$,
the coupling strength $\alpha_X$ relative to $\tilde{\alpha} = 1.7 
{\cdot} 10^{-6}$ (the axion-nucleon coupling strength),
the invariant mass $m_X$, and the references.
Values for B$_X$ and $m_X$ have been derived at 95\,\% CL.}}
\end{table}

On the basis of the available data on the IPC 
anomaly\,\cite{fokke2,fokke3}
an \mbox{$X$-boson} scenario has been proposed, where $X$ stands for an 
isoscalar pseudoscalar particle, with 
its coupling strength ${\alpha}_X$ proportional to the 
isoscalar strength of the M1 transition.  
However, the isoscalar strength in a nuclear transition is in
general unpredictable, because even a relatively small---mostly not well 
known---T=1 component in one of the levels gives a predominant isovector contribution. 
This in turn is because the isovector strength is typically two orders of 
magnitude larger than the isoscalar strength. 
In addition, both the isovector and isoscalar strength range over three orders
of magnitude (see \mbox{Fig.\,5).}
Hence, the sensi\-ti\-vity for detection
of an $X$-boson signal in a given M1 transition from a primarily T=0 level with a 
substantial T=1 component can be estimated only by order of magnitude. 
From Table 1 it can be inferred that the sensitivity of the $X$-boson 
searches---expressed in the ratio of ${\Gamma}_X$ to ${\Gamma}_R$---varies
between 10$^{-8}$ and 10$^{-5}$. 

\vspace{2mm}
{\bf 2. New Experiments}
\vspace{2mm}

In order to test the $X$-boson scenario, a dedicated search
of $e^{+}e^{-}$ angular correlations in isoscalar transitions
is mandatory. The signal observed \cite{fokke2} 
for the relatively pure isoscalar M1 ground state transition from the  
$1^{+}$, $(99.87{\pm}0.12)$\% \cite{adelberger} T=0) level   
at 12.71 MeV in $^{12}$C should be checked
at a high level of significance. For this particular transition the
isoscalar strength can be deduced from ${\Gamma}_{\gamma}$ with 20\,\% 
accuracy  because of the very small (0.0021) T=1 fraction in the level.

To perform such a test, new experiments were carried out
at the Institut f\"{u}r Kernphysik in \mbox{Frankfurt} (IKF) using the proton
and deuteron beams of the 2.5 MV Van de Graaff accelerator. The detector was
the same as the one used in the previous experiments \cite{stiebing,froehlich}.
It consists of eight
${\Delta}$E-E plastic detectors mounted at various angles.
Combinations of these telescopes cover an $e^{+}e^{-}$ correlation angular 
range from 20$^{o}$ to 130$^{o}$, which corresponds to a window of the invariant
mass of 3 to 15 MeV/c$^{2}$ for a neutral boson at a transition energy of 18
MeV for coincidences between electrons and positrons.

For absolute normalisation of the population of relevant nuclear 
levels, a large \mbox{135 cm$^{2}$} Ge(Li) detector
was used as monitor. It was mounted under 90$^{o}$ 
with respect to the beam direction in order to 
minimise Doppler broadening.
The FWHM resolution was ${\sim}20$ keV at 6.13 MeV and ${\sim}30$ keV at 17.64 MeV,
sufficient to separate photo peaks, single escape and 
double escape peaks of the relevant ${\gamma}$-transitions.

As a consistency check of the apparatus and its calibration, the angular 
correlation of 
electron-positron pairs in the $0^{+}{\rightarrow}0^{+}$ E0 decay 
from the 6.05 MeV $0^{+}$ level in $^{16}$O following the 
$^{19}$F($p,{\alpha}e^+e^-$) reaction at 1.6 MeV was remeasured 
\cite{stiebing,froehlich}.  

In addition, the null effect for
the E1 decay from the $1^{-}$ level at 17.23 MeV in $^{12}$C
was verified by remeasuring the $^{11}$B + $p {\rightarrow}$ $^{12}$C 
capture reaction at 1.6 MeV \cite{fokke1}. 

\vspace{2mm}
{\bf 2a. The $^{7}$Li$(p,e^{+}e^{-})^{8}$Be$^*$ reaction at 1.1 MeV}
\vspace{2mm}

In the $1^{+}$ level at 18.15 MeV in $^{8}$Be, the isoscalar component
amounts to 90\% compared to 5\% \cite{barker,paul1}
in the 17.64 MeV level, 
where the anomaly was first observed. 
Assuming that the isoscalar transition strengths T$_{S}$ of the  
transitions to the pure T=0 ground state and the first excited state---with
similar energies and in the same nucleus---scale
with the T=0 component of the depopulating level,
a factor twenty larger anomaly may be expected in the transitions from 
the 18.15 MeV level, compared to what was actually measured \cite{fokke1}
for the 17.64 MeV level. 
 
The angular correlation of the $e^{+}e^{-}$ decay from
the 18.15 MeV resonance at 1.05 MeV was
studied by means of the
$^{7}$Li + p $\rightarrow ^{8}$Be$^{*}$ capture reaction at a
proton energy of 1.1 MeV, for a period of four days. Targets of natural 
Li$_{2}$O with
typical thicknesses of 600 $\mu$g/cm$^{2}$ were used.
From an analysis of the Ge(Li) $\gamma$-spectrum, 
it follows that the 17.64 MeV level is 
populated by 25\% with respect to the excitation of the 18.15 MeV resonance.

\begin{figure}[h]
\begin{center}
\epsfxsize=10.cm
\epsffile{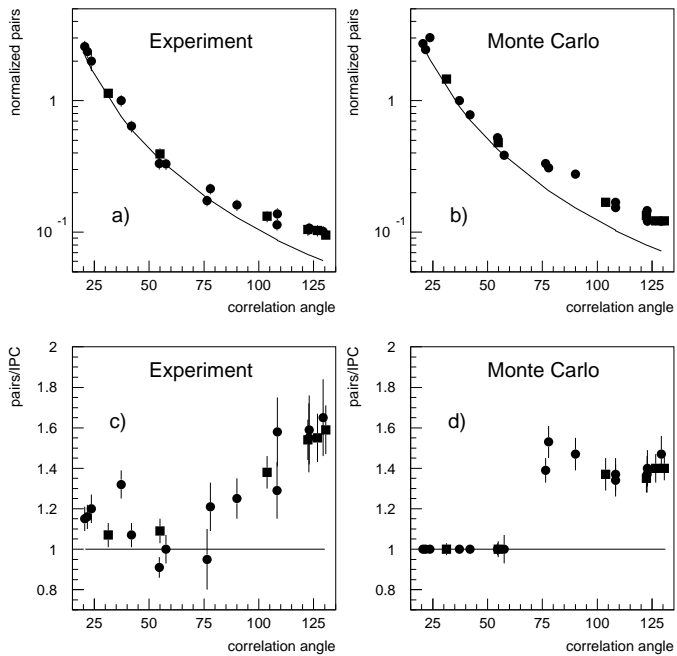}
\end{center}
\caption{\small \em Angular correlation of $e^{+}e^{-}$ pairs following the 
$^{7}$Li$(p,ne^{+}e^{-})^{8}$Be reaction at 1.1 MeV bombarding energy, 
for a) data for the 18.15 MeV resonance in $^{8}$Be with IPC prediction (solid curve);
b) GEANT Monte Carlo simulation \cite{fokke2} of $e^{+}e^{-}$ pairs 
of the 15.15 MeV transition
from the 18.15 MeV level to the first excited state due to IPC (solid curve) 
and additional pairs originating from the two-body decay of a neutral boson 
with a mass of 
9.5 MeV/c$^{2}$ and a branching ratio of $5.8{\cdot}10^{-4}$; 
c) the ratio of the values from above (a) divided by the corresponding 
IPC values; similarly, d) the values of (b) divided by the corresponding IPC 
values. Circular dots denote correlations between the six "small" detectors,
squares between "small" detectors and one "large" detector 
\cite{fokke1,stiebing,froehlich}.}  
\end{figure}

In Fig.\,1a the experimental angular correlation is displayed for 
$e^{+}e^{-}$ pairs integrated over all sum energies of the different
angular combinations. 
Fig.\,1c shows the measured
values divided by IPC. The data are normalised   
to unity for the 
correlation angles at 42$^{o}$, 54.6$^{o}$ and 57$^{o}$. 
At these angles contributions from EPC can be neglected. An excess of pairs 
is observed above $76^{o}$ rising towards the largest angles.
However, this is different from expectations since GEANT MC simulations 
for emission of a 9 MeV/c$^{2}$ boson in the ground state transition  
yield an enhancement 
starting at $55^{o}$. From a comparison with the simulations 
an upper limit of $4.1{\cdot}10^{-4}$ is derived
for the $X$-boson branching ratio in this transition.
We note that in case of a slight boson excess at the normalisation angles,
any possible evidence for a large angle enhancement would decrease.

The excess at large angles can be consistent with $X$-boson emission
in the 15.15 MeV transition depopulating
the 18.15 MeV resonance to the first excited state in $^{8}$Be.
This is illustrated in Figs.\,1b and 1d
 displaying MC simulations
for a 9.5 MeV/c$^{2}$ $X$-boson in the 15.15 MeV transition
with a branching ratio of $(5.8{\pm}2.2){\cdot}10^{-4}$.

\vspace{2mm}
{\bf 2b. The $^{11}$B($d,ne^+e^-)^{12}$C$^*$ reaction at 1.6 MeV}
\vspace{2mm}

In the $X$-boson scenario \cite{fokke2,fokke3} a strong $e^{+}e^{-}$ signal 
\cite{budaThesis,budaNIM,budaphysrev} 
with a branching ratio with respect to the ${\gamma}$-ray intensity of about 
half the IPC coefficient of
$2.7{\cdot}10^{-3}$ can be expected \cite{fokke2,fokke3} for the pure
isoscalar M1 transition from the 12.71 MeV 1$^{+}$, T=0 level to the 
ground state in $^{12}$C. 
The $^{12}$C($p,p')^{12}$C$^*$ and the
$^{11}$B($d,n)^{12}$C$^*$ reactions at subthreshold beam energies
for the 15.11 MeV 1$^{+}$, T=1 level appear to be appropriate to 
investigate the 12.71 MeV resonance.
Using the IKF Van de Graaff accelerator we could  
utilise the $^{11}$B$(d,ne^+e^-)^{12}$C$^{*}$ stripping reaction 
at a deuteron energy of 1.6 MeV leading to the 12.71 MeV $1^{+}$
level with a ${\sigma}_{\gamma}$ of 1.2 mb \cite{kavanagh}. 
Unfortunately the parallel excitation branch $^{11}$B($d,p)^{12}$B forms 
several sources of background: 
\begin{enumerate} 
\itemsep 0.00005ex 
\item The $^{12}$B ground state decays with a ${\beta}^{-}$-spectrum
(endpoint 13.17 MeV) with a cross-section as large as 380 mb \cite{kavanagh}
in 97.1\% of the cases to the $^{12}$C ground state with $t_{\frac{1}{2}}$ = 
20.2 ms. Random ${\beta}^{-}$ coincidences produce a broad
background, with maximum energy at ${\sim}$13 MeV, which---dependent
on beam intensity---can  completely obscure 
the 12.71 MeV sum-energy peak of interest. 
\item The 1.3\% ${\beta}^{-}$-branch 
populating the first excited state at 4.44 MeV followed by internal 
(IPC) \cite{rose} and external (EPC) ${\gamma}$-conversion \cite{borsellino} 
allow real coincidences to produce a 
broad distribution with a maximum around 6 MeV. 
\item The 3.4 MeV $e^+e^-$ 
sum-energy peak of the 4.44 MeV level in
$^{12}$C appears as intrinsic background, being partly directly populated in 
the ($d,n$) reaction.
\end{enumerate}
These background sources were largely suppressed by
implementing a beam pulser into the beam line.
The beam was periodically deflected
from the target into a Faraday cup. The period on target 
and the deflection time could be adjusted over a wide range.
A beam on/off setting of 1.0/100 ms  
was found optimal.

In the production runs the total beam intensity had still to be limited 
to 4 ${\mu}$A, the beam on target 
being 40 nA.  Two targets were used, one consisting of more than 
99.9\% enriched $^{11}$B with a thickness of 600 ${\mu}$g/cm$^{2}$ 
sandwiched between 2 ${\mu}$ thick titanium foil, 
manufactured at JINR (Dubna).
The other was made at KVI (Groningen) and consisted 
of 98\% enriched $^{11}$B evaporated with a thickness of 600 ${\mu}$g/cm$^{2}$ 
on a 2 ${\mu}$ thick gold foil.
In a two weeks period an effective beam time of 100 hours was reached 
with the KVI target in place.  For the Dubna target a shorter beam time
 of 44 hours was left to accumulate which is reflected in the larger error
bars. The occurrence of $p$-capture on $^{11}$B   
from any accelerated molecular hydrogen (contamination in the deuterium gas 
bottle)
was suppressed by a gas stripper in front of the analysing
magnet of the \mbox{Van de Graaff accele\-rator.}     

The population of the 12.71 MeV level in the $^{11}$B(d,n)$^{12}$C reaction 
was obtained by detecting in the Ge(Li) detector the 12.71 MeV and 8.27 MeV
${\gamma}$-rays to the $^{12}$C ground state and
first excited state. 
In the ${\gamma}$-ray spectra from both targets a peak
occurs at 7.64 MeV from the decay of the degenerate 3$^{+}$ and 2$^{+}$ 
levels at ${\sim}$9.35 MeV in $^{28}$Si. These levels are selectively 
populated in the
$^{27}$Al($d,n$) reaction \cite{lawergren}, presumably 
due to (beam halo) deuterons reacting with the aluminium 
target frames. 
Moreover, a contamination of 2\% $^{10}$B in the KVI target inducing 
the $^{10}$B$(d,p)^{11}$B reaction [22-24] 
caused a strong contribution of ${\gamma}$-rays at 4.49, 6.76 and 8.92 MeV. 

Random coincidences due to the remaining ${\beta}^{-}$ background were 
subtracted by selecting windows on the prompt and delayed parts of the time
spectra. In the sum-energy spectra obtained in this way for the 
different runs, a complex of several sum peaks could be identified at 
21.8$^{o}$, one of the smallest correlation angles between neighbouring
detectors, which contain the highest statistics but also a substantial EPC
contribution, and at the correlation angles 37.2$^{o}$, 42.0$^{o}$ and 
54.6$^{o}$ (see Fig.\,2). Sum-energy lines corresponding with the 
observed ${\gamma}$-ray transitions are indicated.
A comparison of the widths of the 4.44 MeV sum-energy lines in the KVI and 
Dubna spectra shows that the 2 ${\mu}$ supporting titanium layer does 
not notably influence the 
resolution in the latter spectrum. 

\begin{figure}[ht]
\begin{center}
\leavevmode
\epsfxsize=12.cm
\epsffile{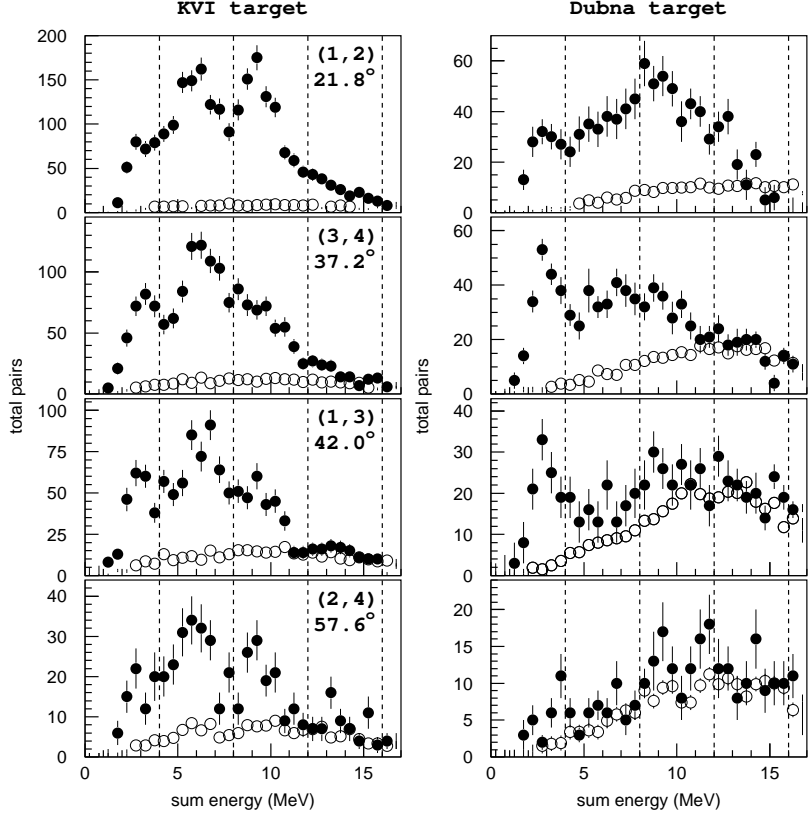}
\end{center}
\caption{\small \em Sum-energy spectra for $e^+e^-$ pairs following 
(predominantly) the $^{11}$B$(d,n)^{12}$C
reaction at 1.65 MeV using KVI and Dubna targets for correlation angles at 
21.8 $^{o}$, 37.0 $^{o}$, 42.0 $^{o}$ and 57.6 $^{o}$. The solid respectively 
open circles represent the total coincidence yield and the yield of 
accidental coincidences.
A 4 MeV wide binning used in Fig.\,3 is indicated with solid vertical lines.}
\end{figure}

A global analysis of the spectra has been made by integrating the yields
of the spectra in four 4 MeV wide intervals without attempting to fit the
complex structures. In Fig.\,3 from top to bottom the background
subtracted angular 
correlations are exhibit for the two targets. 
The curves show parameterised
functions \cite{froehlich,goldring} representing the theoretical \cite{rose}
IPC angular distributions of M1 transitions. 
Like for the $^{8}$Be experiment, they are normalised to unity for the 
correlation angles at 42$^{o}$, 54.6$^{o}$ and 57$^{o}$, 
where EPC is not expected to contribute significantly. 
\begin{figure}[ht]
\begin{center}
\leavevmode
\epsfxsize=12.cm
\epsffile{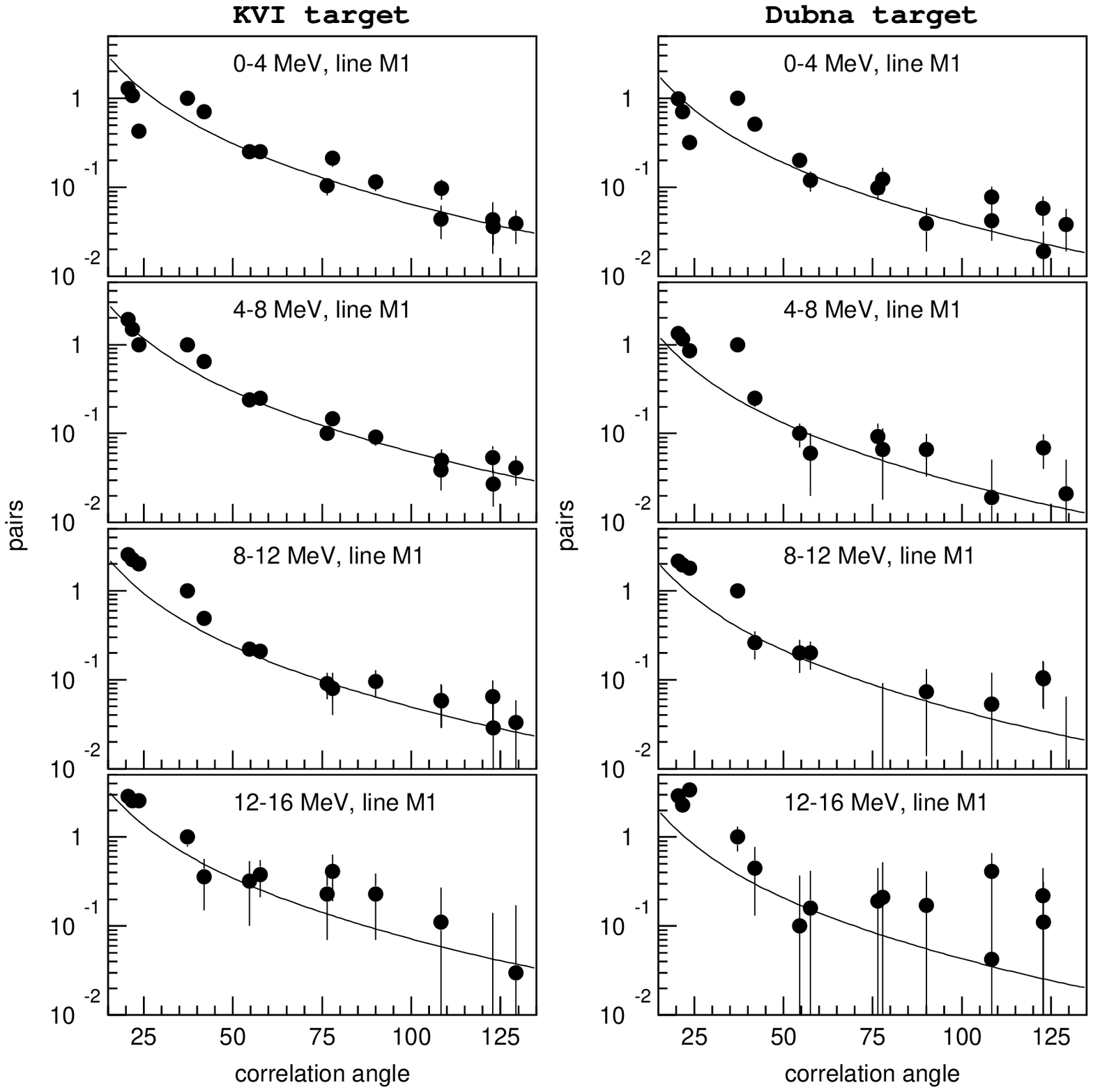}
\end{center}
\caption{\small \em Angular correlations for $e^+e^-$ pairs with sum-energy 
intervals 0-4 MeV, 4-8 MeV, 8-12 MeV and 12-16 MeV, produced in 
the $^{11}$B$(d,n)^{12}$C
reaction at 1.65 MeV using two different targets (left: KVI, right: Dubna). 
The curves are parameterised functions\cite{froehlich,goldring} of theoretical 
IPC distributions for M1 transitions fitted 
through the data.} 
\end{figure}
A fair agreement with IPC assuming M1 character
is observed over the whole angular range for $e^+e^-$ pairs
in the first two intervals with sum-energies below 8 MeV.
Deviations from IPC show up for the \mbox{8 - 12 MeV} and 
\mbox{12 - 16 MeV} intervals. 

\begin{figure}[ht]
\begin{center}
\vspace{-2cm}
\leavevmode
\epsfxsize=12.cm
\epsffile{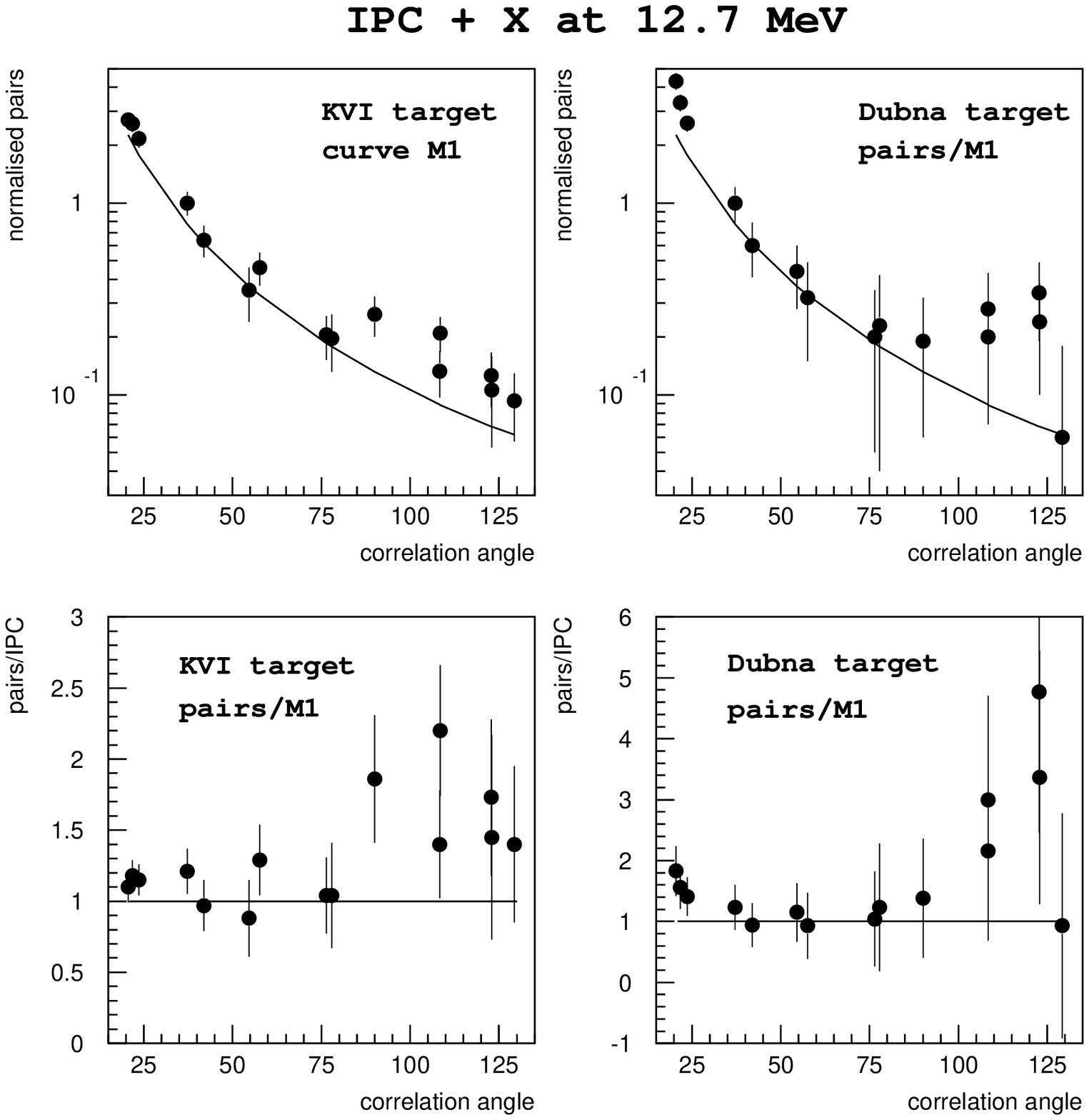}
\end{center}
\caption {\small \em Top: Angular correlations 
for sum-energy peaks from internal pairs at 12.71 MeV
produced in the $^{11}$B$(d,n)^{12}$C
reaction at 1.65 MeV, using the two targets (KVI, Dubna). 
The curves represent theoretical IPC values for M1 transitions fitted 
through the data. Bottom: The experimental values (from the top) divided by 
the theoretical IPC distributions.}
\end{figure}

The correlation of the $e^+e^-$ pairs in a 4 MeV window around the 
sum-energy peak of the 12.71 MeV transition is 
shown in the top figures of Fig.\,4. 
A clearly visible but statistically marginal enhancement over 
IPC shows up at large correlation angles ${\geq}90^{o}$ for both the KVI
and the Dubna target. In the bottom figures these distributions are 
divided by the corresponding IPC values. They are normalised to unity for
the correlation angles between 40$^{o}$ and 60$^{o}$.  
The combined statistical significance amounts to 3.5${\sigma}$. 

A comparison with the simulations for the  
two-body decay of a short-lived neutral boson yields an invariant mass of
($9 {\pm} 1$) MeV/c$^{2}$ and a branching ratio of
$(7{\pm}3){\cdot}10^{-4}$, a factor two below---though within $1{\sigma}$
compatible with---the value of $(1.6{\pm}0.7){\cdot}10^{-3}$ given 
in \cite{fokke2,fokke3}. An $\alpha_X$ value of $18 \pm 7$ has been 
derived with Eq.\,1 and listed in Table\,1.
The fair agreement of this enhancement with the previous observation 
\cite{fokke2,fokke3} motivates 
a high statistics experiment\,\cite{fokke4} in the future.

\vspace{2mm}
{\bf 3. Discussion of the results}
\vspace{2mm}

In light nuclei the experimental ${\gamma}$-ray transition width 
${\Gamma}_{\gamma}$ from isospin mixed levels can be expressed as the sum of 
the scalar and isovector transition width ${\Gamma}_S$ and 
${\Gamma}_V$. 
The transition strength T$_{\gamma}$ is defined as the transition width 
${\Gamma}_{\gamma}$ divided by the \mbox{Weisskopf} estimate or 
unit (${\Gamma}_W$ or w.u).
For transitions to pure T=0 states, it is the sum of the separate isoscalar
and isovector strength T$_S$ and T$_V$, scaled by the T=0 (T=1) fraction 
$x$ (1-$x$) in the depopulating level, according to: 
\begin{equation}
{\rm T_{\gamma}} = \frac{{\Gamma}_{\gamma}}{{\Gamma}_W} = 
\frac{{\Gamma}_S}{{\Gamma}_W} + 
\frac{{\Gamma}_V}{{\Gamma}_W} =
 {\it x}{\rm T_S} + (1-{\it x}){\rm T_V}.
\end{equation}

\begin{table}[h]
\scalebox{0.85}{
\begin{tabular}{|ccccccccc|}
\hline
E$_{R}$&
T=0 &
E$_{\gamma}$&
${\Gamma}_{\gamma}$&
${\Gamma}_{W}$ & 
${\Gamma}_{S}$ & 
${\Gamma}_{V}$ & 
T$_{S}$ &
T$_{V}$ \\
 MeV&
$x$
&MeV& eV & eV &eV   & eV& &        \\
\hline
12.71&$0.9979{\pm}0.0012$&    12.71 & $0.35{\pm}0.05$ & 43.2 & 
 $0.30{\pm}0.04$ &  $0.05{\pm}0.04$ &
$(7.0{\pm}1.3){\cdot}10^{-3}$&
 $0.39{\pm}0.28$ \\
          15.11& $0.0021{\pm}0.0012$&    15.11 & $38.5{\pm}0.8$ & 70.8 &   
 - &  $38.5{\pm}0.8$ &
 - & $0.54{\pm}0.01$ \\
 17.64& $0.05{\pm}0.01 $&17.64 &$16.7{\pm}0.8$  & 113.0 & 
 $1.0{\pm}0.4$&  $15.7{\pm}0.8$ &
 $0.18{\pm}0.09$
& $0.15{\pm}0.02$ \\
      &   & 14.64 & $8.2{\pm}0.4$ & 64.5 &
 $0.5{\pm}0.2$ &  $7.6{\pm}0.5$ &
$0.15{\pm}0.07$   & $0.12{\pm}0.01$       \\
          18.15& $ 0.90{\pm}0.05$ & 18.15 & $3.0{\pm}0.3$   & 123.0 & 
 ${\leq}0.95$&  $2.7{\pm}0.3$ &
   ${\leq}8.6{\cdot}10^{-3}$  
& $0.22{\pm}0.12$ \\
                &          & 15.15 & $3.8{\pm}0.4$   & 71.5& 
 $1.6{\pm}0.9$&  $2.2{\pm}1.0$ 
& $(2.5{\pm}1.4){\cdot}10^{-2}$ 
    &  $0.30{\pm}0.14$        \\
\hline
\end{tabular}
}
\caption{\small \em Parameters of ${\gamma}$-decay properties for M1 
transitions depopulating $1^{+}$ \mbox{levels} in $^{12}$C and $^{8}$Be.
Listed are the resonance energy $E_{R}$, the T=0 fraction of
the level expressed in the parameter $x$, the transition energy $E_{\gamma}$,
the $\gamma$-width $\Gamma_{\gamma}$, the Weisskopf estimate $\Gamma_W$,
the isoscalar and isovector partial widths $\Gamma_S$ and $\Gamma_V$, 
and the corresponding transition strengths $T_{S}$ and $T_{V}$.} 
\end{table}

A compilation of known transitions strengths T$_V$ and T$_S$ for pure 
isovector (${\Delta}$T=1, $x$=0) respectively isoscalar (${\Delta}$T=0,
$x$=1) transitions has been made for self-conjugate (A = 2Z) nuclei by 
Skorka et al.\cite{skorka}.
Histograms of the logarithm of the strength for both cases are shown 
in Figs.\,5a and 5b.  
We assume that for unmeasured transitions the probability of the isoscalar
respectively isovector strength is according to these distributions.

\begin{figure}[h]
\begin{center}
\epsfxsize=10.cm
\epsffile{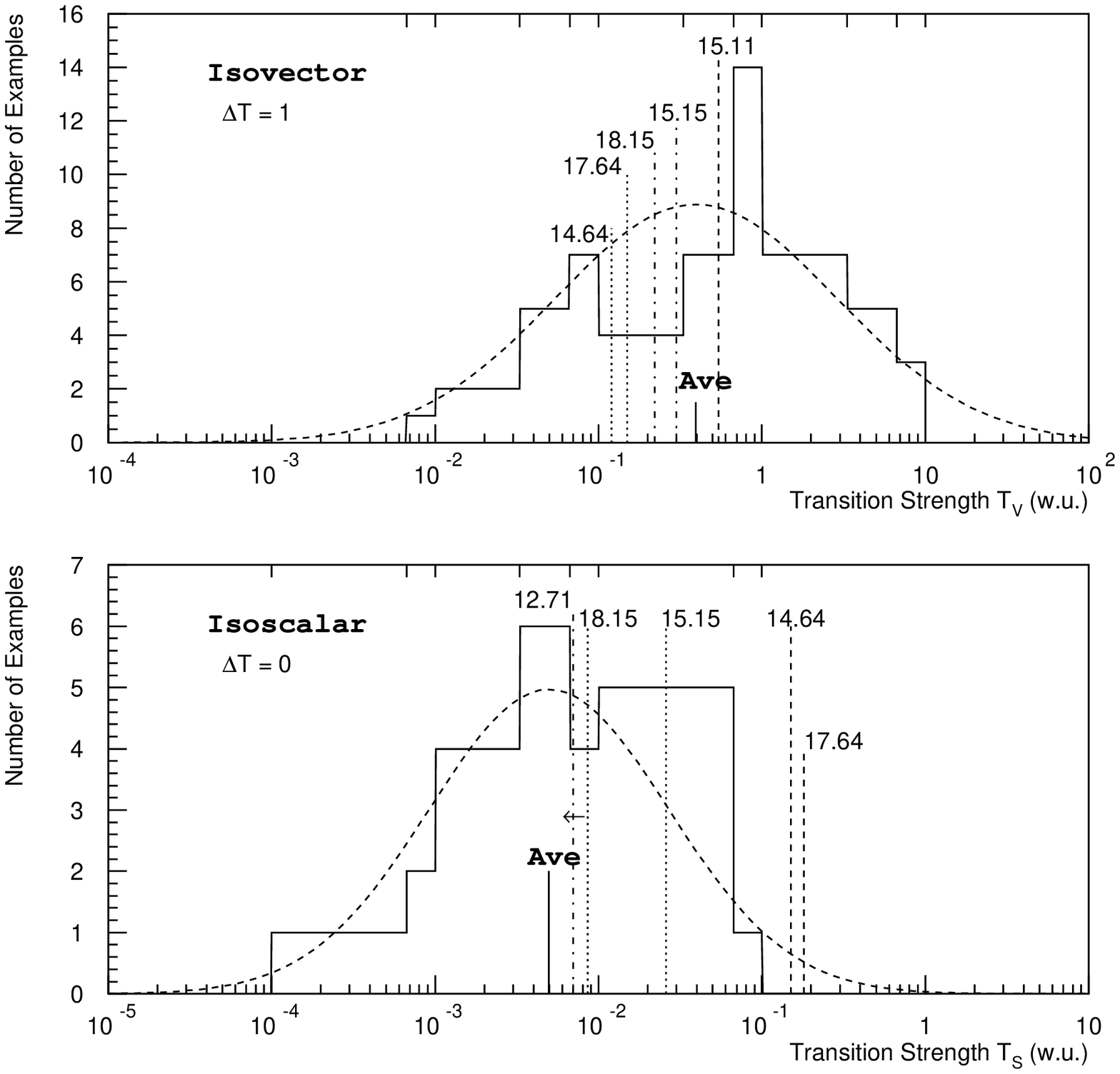}
\end{center}
\caption{\small \em A histogram for measured examples of isovector (top) and 
isoscalar (bottom)
transition strength expressed
in Weisskopf units (w.u) on a log-scale.
Dashed lines are Gaussian fits through the histograms. Straight 
vertical lines show the average values denoted by {\bf Ave}. 
Vertical lines represent T$_V$ and 
T$_S$ values, derived using equation 3,
for the transitions presently under investigation.}
\end{figure}
Both (on a log-scale) Gaussian shaped distributions have a FWHM corresponding to two
orders of magnitude, and standard deviations of factors 7.4 and 10.6
respectively.  
The average values for the isovector respectively isoscalar transitions  
were found \cite{skorka} as T$_V^{ave} = 0.395$ and 
T$_S^{ave} = 4.97{\cdot}10^{-3}$. 
They are indicated in Fig.\,5. 
The ratio 
T$_S^{ave}$/T$_V^{ave}$
 is 0.0126, in fair agreement 
with the value of 0.0082 given by de Shalit and Fesh\-bach \cite{shalit} in 
the long wave length limit.  

The levels at 12.71 and 15.11 MeV in $^{12}$C have nearly pure T=0 respectively
T=1 character\,\cite{adelberger}.
A T$_S$ value for the 12.71 MeV ground state transition has been derived 
using Eq.\,2 while taking the average T$_V$ value\,\cite{skorka} with an 
error such as to include all T$_V$ values of Table\,2. 
The small error in T$_V$ for the 15.11 MeV transition reflects the 
small value of $x$ in Eq.\,2.   
This specific T$_S$/T$_V$ ratio, 0.0129 ${\pm}$ 0.0024, is 
in good agreement with T$_S^{ave}$/T$_V^{ave}$ and also
with the theoretical expectation. The isospin structure of these essentially
pure transitions is thus sufficiently well understood to be used for 
further analysis.

For consistency evaluation the $X$-boson scenario can be tried out as 
{\em analyser} of isoscalar width and consequently strength.
Using Eq.\,1 the isoscalar and isovector transition widths 
${\Gamma}_S$ and ${\Gamma}_V$ can straightforwardly be derived as 
\begin{equation}
{\Gamma}_S =  {\Gamma}_{\gamma} 
 (\frac{ {\alpha}_X } { {\alpha}_X^{o} }), \ \
{\Gamma}_V = {\Gamma}_{\gamma}
[1 - (\frac{{\alpha}_X}{{\alpha}_X^{o}})].
\end{equation}
From the average ${\alpha}_X$ value of $21 {\pm} 6$ 
for the 12.71 MeV ground state transition (see Table 1)  
a maximum boson-nucleon coupling strength ${\alpha}_X^{o}=24 {\pm} 8$ 
can be expected for a pure isoscalar transition.   
The width parameters ${\Gamma}_S$ and ${\Gamma}_V$ 
for the M1 transitions under investigation calculated using Eq.\,3 are
listed in Table 2.
The related strength parameters T$_S$ and T$_V$, 
taking the old literature isospin assignments\,\cite{barker,paul1}, 
are included in Table 2 and in Fig\,5.

The T$_{V}$ values are (on the log-scale) all within 
$0.5{\sigma}$ in the middle of the distribution in Fig\,5\,(top). 
In Fig\,5 (bottom) the T$_S$ values are more scattered: those for 
the 17.64 and 14.64 MeV are at the tail ($1.5{\sigma}$)
of the distribution. Within the typical errors in ${\Gamma}_S$
(${\sim}$50\%, see Table\,2), they are factors 36 and 30 
enhanced with respect to the average T$_S$ value (Ave).

The deduced T$_S$ values for the transitions from the 18.15 MeV level
are within 0.5${\sigma}$ consistent with the average T$_S$ value. 
The relative smallness of the upper limit for the T$_S$ value
of the 18.15 MeV transition---about a factor 20 below the T$_S$ values
for the 17.64 decay---could explain the absence of a visible 
anomaly. 
To test the scenario in this case
would require at least an order of magnitude higher accuracy in the 
measurement, which is at present difficult to attain.

\vspace{2mm}
{\bf 4. Summary}
\vspace{2mm}

Results of two dedicated experiments are reported yielding  
further indications for an anomaly at 9 MeV/c$^{2}$
in the angular correlation of IPC.  
The first experiment ($^8$Be) shows a deviation from IPC at large
correlation angles presumably due to the same anomaly in the
 transition to the first excited state. 
The second experiment ($^{12}$C) shows a relatively large anomaly at 
9 MeV/c$^{2}$ albeit with limited statistics.
Both results are compatible with an $X$-boson scenario where 
the boson-nucleon coupling 
strength is proportional to the isoscalar strength in the M1 transition.
Exploiting isospin structure as a guideline, further high statistics 
experiments are needed to establish the nature of the anomaly. 

\vspace{2mm}
{\bf Acknowledgement}
\vspace{2mm}

We would like to thank the staff of the 2.5 MV Van de Graaff accelerator
at IKF for providing clean and stable proton and deuteron beams.
This work has been supported by the Bundesministerium f\"{u}r
Forschung and Tech\-no\-logy of the Federal Republic of Germany.
Some of us (RvD, JvK) have been supported by the Dutch research organisations  
FOM and NWO. One of us \mbox{(FWNdB)} gratefully acknow\-ledges support
from Promis Engineering Advisory Agency, The Netherlands. 


\end{document}